\documentclass[runningheads]{llncs}
\usepackage{color,soul}

\usepackage{enumitem}
\usepackage[dvipsnames]{xcolor}
\usepackage{graphicx}
\graphicspath{ {./images/} }
\usepackage[cmex10]{amsmath}

\usepackage{algorithm}
\usepackage[noend]{algcompatible}
\usepackage[nodisplayskipstretch]{setspace}
 \usepackage[font=scriptsize,labelfont=bf]{caption}
\begin{document}
%Provisioning Energy Services for Quality of Experience\\
\title{
Maximizing Consumer Satisfaction \\ of IoT Energy Services}

% \author{Amani Abusafia\orcidID{0000-0001-9159-6214} \and
% Athman Bouguettaya\orcidID{0000-0003-1254-8092}}
% %
% %\authorrunning{F. Author et al.}
% \institute{The University of Sydney, Sydney, NSW 2000, Australia
% \email{\{amani.abusafia,athman.bouguettaya\}@sydney.edu.au}}
% %

\author{ Amani Abusafia\and
Athman Bouguettaya \and
Abdallah Lakhdari}
\authorrunning{A. Abusafia et al.}
%{alak5184,athman.bouguettaya,azadeh.gharineiat}@sydney.edu.au
\institute{The University of Sydney, Sydney NSW 2000, Australia\\
%\email{\{alak5184,athman.bouguettaya,azadeh.gharineiat\}@sydney.edu.au}\\}
\email{\{amani.abusafia,athman.bouguettaya,abdallah.lakhdari\}@sydney.edu.au}}

\maketitle              % typeset the header of the contribution
\vspace*{-15pt}
\begin{abstract}
%We propose a novel \textit{incentive-based} framework for composing energy service requests. An incentive model is designed that considers the context of the providers and consumers to determine \textit{rewards} for sharing wireless energy. We propose a novel priority scheduling approach to compose energy service requests that maximizes the reward of the provider. A set of exhaustive experiments with a dataset and collected IoT users' behavior is conducted to evaluate the proposed approach. Experimental results prove the efficiency of the proposed approach.  
 %A novel QoE model is designed as a criterion to optimize the composition of  energy services.
We propose a novel \textit{Quality of Experience (QoE)}-aware framework to crowdsource IoT energy services efficiently. The proposed framework leverages the provisioning of energy services as an auxiliary to increase consumers' \textit{satisfaction}. A novel QoE model is developed as a metric to assess the consumers' satisfaction with the provisioning of energy services. Two novel composition algorithms, namely, Partial-Based (PB) and Demand-Based (DB) approaches, are proposed to ensure the highest QoE for consumers. Both approaches leverage the providers' \textit{flexibility} and  \textit{shareable} nature of energy services to efficiently allocate services and optimize the QoE.   A set of extensive experiments is conducted to evaluate the proposed approaches' efficiency and effectiveness.\looseness=-1


\end{abstract}

\section{Introduction}

% IoT SERVICES
% crowdsourcing
% energy sharing
% benefits
% QoE 
% environment

% QoE in general
% IoT services
% Energy service
% \textit{Internet of Things (IoT)} is an emerging technology that aims to connect every tangible device or \textit{thing} to the internet. The number of connected devices is expected to reach 125 billion in 2030 \cite{markit2017internet}.  IoT devices, such as wearables, are usually equipped with capabilities, including identifying, sensing, networking, and processing  \cite{whitmore2015internet}. The pervasiveness of IoT devices offers opportunities to abstract their capabilities as \textit{IoT Services}. IoT services refer to services provided by IoT devices to other nearby IoT devices. Examples of IoT services are WiFi hotspots \cite{bahutair2019adaptive}, processing power services \cite{habak2015femto}, and energy services \cite{lakhdari2020Vision}\cite{lakhdari2018crowdsourcing}.

\textit{Internet of Things (IoT)} is a paradigm that enables everyday objects (i.e., \emph{things}) to connect to the internet and exchange data. IoT devices, such as smartphones and wearables, usually have augmented capabilities including sensing, networking, and processing \cite{whitmore2015internet}. Abstracting the capabilities of these IoT devices using the \emph{service paradigm} may yield to  multitude of novel \emph{IoT services} \cite{lakhdari2021fairness}. These IoT services may be \textit{exchanged} between {IoT devices} as \textit{crowdsourced} IoT services. For example, an IoT device may offer WiFi hotspots or wireless energy services to charge other IoT devices \cite{lakhdari2021fairness}. These crowdsourced IoT services present a convenient and cost-effective solutions \cite{lakhdari2021fairness}. Our focus is on wireless energy sharing services among IoT devices.\looseness=-1

%The  augmented capabilities of IoT devices offers opportunities to \textit{crowdsource} them as \textit{IoT Services}. 

\emph{Energy-as-a-Service (EaaS)} is the abstraction of the wireless delivery of energy among nearby IoT devices \cite{lakhdari2018crowdsourcing}\cite{lakhdari2021fairness}. \textit{EaaS} is an IoT service where energy is delivered from an energy provider (e.g., a smart shoe or smartphone) to an energy consumer (e.g., a smartphone) through wireless means. % (see Fig.\ref{fig:scenario} (B)). Smart textiles or smart shoes are examples of energy providers which may \textit{harvest} energy from natural resources (e.g., body heat or physical activity) \cite{kwak2019textile}\cite{gorlatova2015movers}. 
%Wearables such as smart textiles or smart shoes may \textit{harvest} energy from natural resources, e.g. body heat or kinetic activity \cite{kwak2019textile}\cite{gorlatova2015movers}. For example, wearing a PowerWalk harvester may generate energy to charge four smartphones from an hour walk at a comfortable speed\footnote{bionic-power.com}. The harvested energy may be offered to close-by IoT devices as a service. 
\textit{EaaS} may be deployed through the newly developed ``Over-the-Air" wireless charging technologies \cite{lakhdari2020composing}\cite{OvertheAirCharger}. Several companies, including Xiaomi\footnote{mi.com}, Energous\footnote{energous.com}, and ossia\footnote{ossia.com}, are currently developing wireless charging technologies for IoT devices over a \textit{distance}. For example,  Energous developed a device that can charge up to 3 Watts power within a 5-meter distance.\looseness=-1

The crowdsourced EaaS ecosystem is a \textit{dynamic} environment that consists of \textit{providers} and \textit{consumers} congregating in \textit{microcells}. A microcell is any confined area where people may gather (e.g., coffee shops). % (see Fig.\ref{fig:scenario} (A)).
In this ecosystem, IoT devices may share energy with nearby IoT devices. %(see Fig.\ref{fig:scenario} (B)). 
A key aspect to unlocking the full potential of the EaaS ecosystem is to design an \textit{end-to-end} Service Oriented Architecture (SOA) to share crowdsourced energy. We identify three key components of the SOA: energy service \textit{provider}, energy service \textit{consumer}, and \textit{super-provider}. %(see Fig.\ref{fig:SOA}). 
In this architecture, providers advertise services, consumers submit requests, and  super-provider (i.e., microcell's owner) manage the exchange of energy services between providers and consumers. This paper focuses on \textit{managing energy sharing from the super-provider perspective}.\looseness=-1

Super-provider typically focus on ensuring that customers keep coming back to their businesses. Their  revenue is usually directly related to \textit{foot traffic} \cite{muller1994expanded}. Customer \textit{satisfaction} is therefore paramount as a strategy to either maintain or increase the business target revenue \cite{chao2013c}. A key objective is to ensure that customers have the \textit{best experience}. We propose to use energy sharing as a key ingredient to provide customers with the best quality of experience when visiting the business. For example, a case study showed that “Sacred”, a cafe in London, had a noticeable increase in foot traffic after installing wireless charging points\footnote{air-charge.com}.\looseness=-1

We define a \textit{Quality of Experience (QoE)} metric to represent the level of \textit{satisfaction} across energy consumers over a period of time in a specific microcell. Note that QoE is different from Quality of Service (QoS).  QoE uses QoS as a base to express satisfaction of a service over a period of time. {QoE has traditionally been used in domains that assess how users perceive a service \cite{fizza2021qoe}\cite{moller2014quality}\cite{wang2019kaleidoscope}. Our proposed environment requires the use of a different type of QoE. In particular, we identify the following three aspects that shape the new QoE definition: (1) crowdsourced environment resources are usually limited and cannot fulfill all consumers' requirements. Hence, assessing consumers’ satisfaction should consider the limited available resources. Energy services may be provided partially due to the limited resources and the shareable nature of energy services, e.g., a single service may be split into smaller services and provided among multiple consumers. In a limited resource environment, consumers’ experience with partial services differs from complete services. (3) Consumers' satisfaction with energy services will indirectly impact their experience with the super-provider's microcell.} %In a nutshell, our focus is on the experience of consumers within the microcell, not only on the service itself.
Therefore,our research focuses on the \textit{ super-provider's perspective} of QoE.\looseness=-1
%Existing research has mainly focused on the consumer perspective of QoE \cite{lemon2016understanding,wang2019kaleidoscope}. In contrast, our research focuses on the \textit{ super-provider's perspective} of QoE.

%Existing research focuses on assessing either the quality of services \cite{lakhdari2020Vision,lakhdari2021fairness} or the quality of experience   from a consumer perspective. In this paper, we focus on determining the QoE from \textit{a super-provider's perspective}.  
Assessing the QoE from a super-provider's perspective usually entails measuring the \textit{aggregated} satisfaction of consumers over time. \textit{Consumer's satisfaction} is defined as \textit{meeting or exceeding} a set of expected service goals \cite{farris2010marketing}. In this context, we define consumer satisfaction as \textit{receiving the requested energy or part of it}. We focus on  \textit{optimizing} the \textit{QoE} by efficiently provisioning and fulfilling the consumers' energy requirements.

The limited availability of energy is a key challenge that may hinder the super-provider from optimizing the consumers' QoE \cite{lakhdari2021fairness}. For instance, an energy consumer might not find their requested energy at a certain time in the microcell, resulting in an unsatisfying experience. In this context, using traditional resource allocation algorithms may incur uneven energy sharing for some consumers. Therefore, we propose a QoE-driven service provisioning framework to satisfy energy consumers in a crowdsourced IoT environment. The framework requires prior knowledge of providers' temporal preferences and the microcell energy demands. The proposed framework leverages the \textit{shareable} nature of energy services to split the energy between consumers if the required energy is more than the available energy\cite{lakhdari2018crowdsourcing}. Intuitively, the super-provider may prefer to offer part of the required services to all consumers than offering it to some of them. Hence, we propose a heuristic Partial-Based (PB) approach which splits services among consumers in the case of low energy availability. Another possible solution is to leverage \textit{flexible} providers that offer services on \textit{multiple} time slots by allocating their services to the most {demanding} slots. Intuitively, this may ensure a better distribution of the available services. 
%Intuitively, the high-demanding time slots will require more services. Thus, services should be assigned to them prior to less demanding time slots, which may ensure a better distribution of the available services.
%the high-demanding time slots will require more energy, services should be assigned to them first. This may ensure a better distribution of the available services.
Therefore, we additionally propose a heuristic Demand-Based (DB) approach. The DB approach extends the PB approach by prioritizing the allocation of services based on the highest demanding time slots. {The main contributions of this paper are:}

\begin{itemize} [ noitemsep,nosep,leftmargin=10pt,labelsep=1pt,itemindent=0pt, labelwidth=*]
    %\item A Quality of Experience (QoE) model to gauge energy consumers' satisfaction.
    \item A novel Quality of Experience (QoE) model for crowdsourced energy services.
    %\item A Partial-Based QoE-driven composition of energy services.
    %\item A Demand-Based QoE-driven composition of energy services.
    %\item A QoE-aware framework for composing energy services.
    \item A framework for QoE-driven composition of  IoT energy services.
    \item An experimental analysis with two implementations of the proposed QoE-driven energy composition framework.
\end{itemize}

\begin{figure}[!t]
    \centering
        \setlength{\abovecaptionskip}{2 pt}
\setlength{\belowcaptionskip}{-20 pt}
    \includegraphics[width=0.7\linewidth]{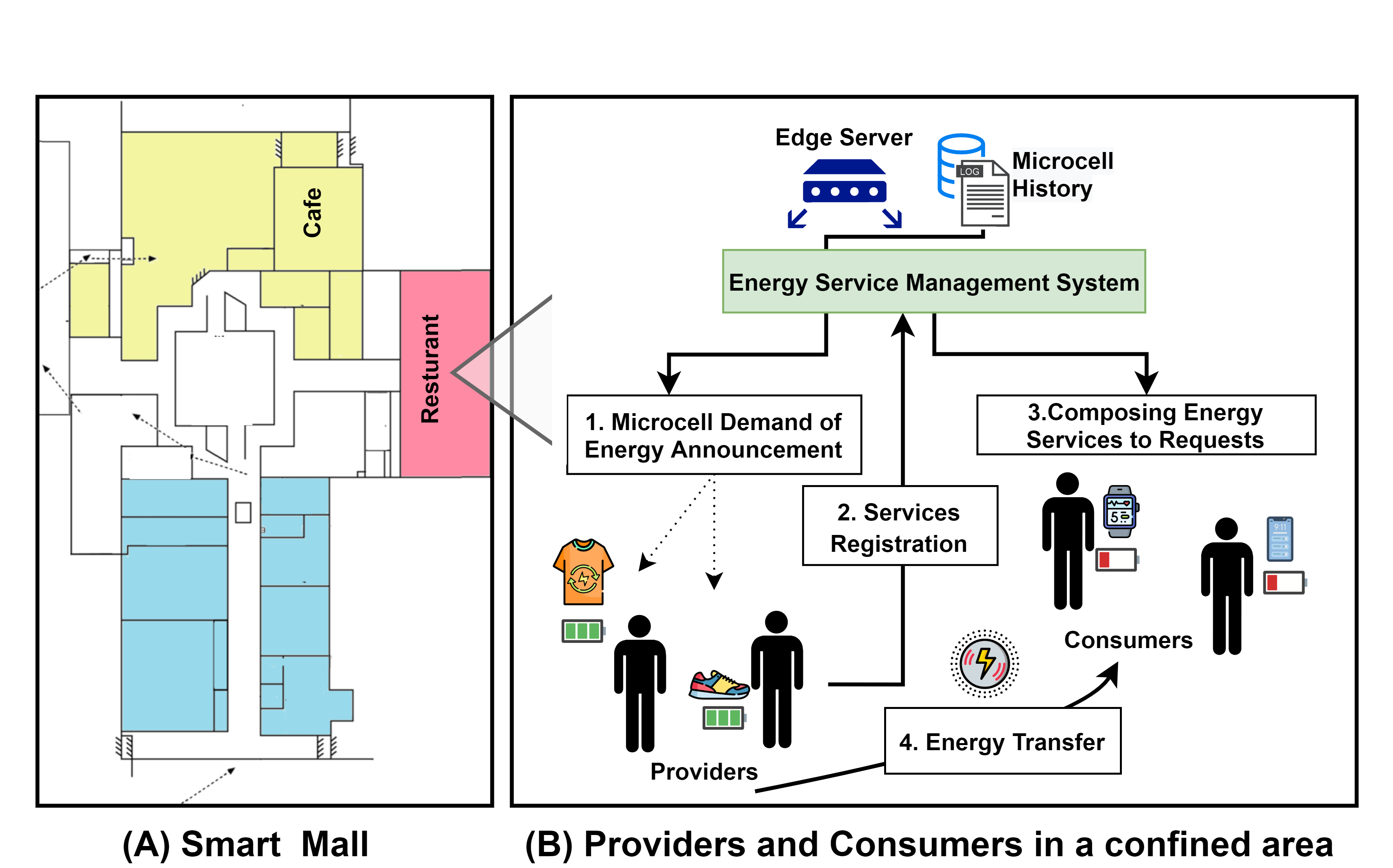}
    \caption{IoT energy services environment}
    \label{fig:scenario}

\end{figure}
\vspace{-15pt}
\subsection{Motivating Scenario}
\begin{figure}[!t]
    \centering
       \setlength{\abovecaptionskip}{1pt}
\setlength{\belowcaptionskip}{-20pt}
    \includegraphics[width=0.9\linewidth]{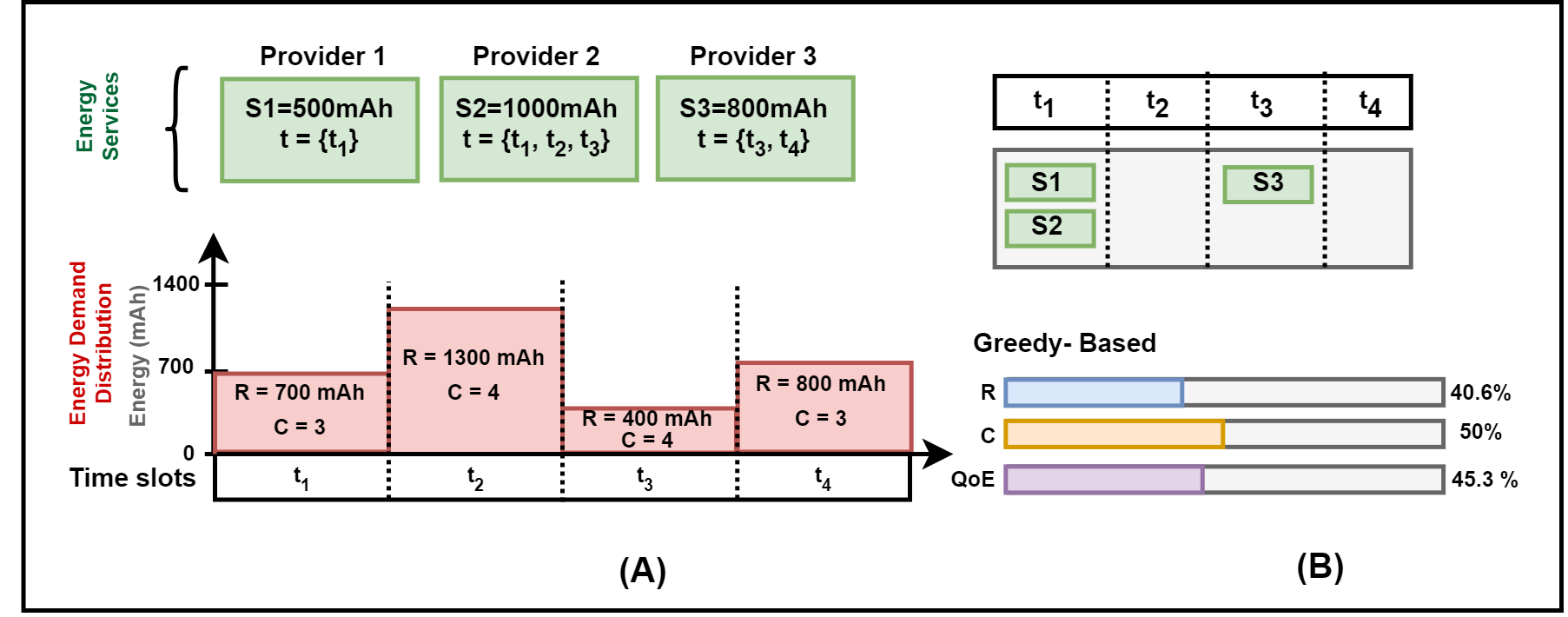}
    \caption{(A) Microcell energy demand and providers services (B) Greedy energy provisioning approach }
    \label{fig:fullscenario}
\end{figure}

\vspace{-5pt}
We describe a scenario in a confined place (i.e., microcell) where people congregate, e.g., cafes,and restaurants (see Fig.\ref{fig:scenario} (A)). Each microcell may have several IoT devices acting as energy providers or consumers (see Fig.\ref{fig:scenario} (B)). The super-provider aims to leverage the crowdsourced energy services as a tool to enhance the consumers' \textit{experience}. We assume all local energy services and requests are submitted and managed at the \textit{edge}, e.g., a router in the microcell (see Fig.\ref{fig:scenario}(B)). {We assume that the super-provider offers incentives to encourage energy sharing in the form of credits. These would be used to receive more energy when the providers act as consumers in the future \cite{lakhdari2021fairness}}   We assume the {super-provider} has a prior knowledge of the \textit{Microcell Energy Demand} ($\mathcal{MED}$) in the microcell over a period of time (T) (see Fig.\ref{fig:fullscenario} (A)). {The $\mathcal{MED}$ may be estimated based on previous history \cite{lakhdari2021proactive}.}  The $\mathcal{MED}$ is represented in terms of the requested energy in each time slot, e.g., 700 mAh at time slot $t_1$. {The granularity of the time slots can also be estimated based on the previous history of the microcell \cite{lakhdari2021fairness}}.\looseness=-1

 %We consider the following scenario to illustrate and motivate our work as shown in Fig.\ref{fig:scenario}.  Assume that a geographical region is split into microcells where people gather, such as cafes, restaurants, etc. Each microcell is also assumed to contain several IoT devices, which act as service providers or consumers (see Fig.\ref{fig:scenario}B). All local energy services and requests are submitted and managed by an \textit{IoT coordinator} at the \textit{edge}, e.g., a router in the microcell. The IoT coordinator uses \textit{rewards} to encourage providers to share energy. Rewards may come in the form of stored credits to providers.  A provider receives a reward based on an incentive model \cite{abusafia2020incentive}. Assume the microcell owner wants to increase the {foot traffic }in the microcell for revenue purposes. The aim is to increase the traffic by enhancing the \textit{quality of experience} for visitors so they are motivated to come back. The quality of experience, in this context, is defined by fulfilling as many energy requests as possible. We assume the IoT coordinator has an \textit{aggregated} microcell energy demand predicted from the\textit{ microcell history}. 
 %The IoT coordinator records the\textit{ microcell history} to predict the \textit{aggregated} microcell demand for energy. 
 
 We also assume that the super-provider has prior knowledge of the providers preferences in terms of time and energy service attributes.  {An incentive model is employed to  predict the amount of energy that would be available for consumption \cite{abusafia2020incentive}}. For instance, provider 1 in Fig.\ref{fig:fullscenario} (A) wants to offer the energy service $S1$ with 500 mAh at time $t_1$. Another example, provider 2  wants to offer $S2$ at time slots $t_1$ , $t_2$, or $t_3$. We assume the provider would stay for the full-time slot. We also assume that the provider's service amount is fixed and can be split among multiple time slots. For instance, provider 2 may share part of their service $S2$ on $t_1$, e.g., 300 mAh, and the other part at $t_2$ or $t_3$. We also assume a single energy provider may share their spare energy with multiple energy consumers, within \emph{a specific time interval}. The super-provider uses \textit{rewards} to encourage providers to share energy. Rewards may come in the form of stored credits to providers. A provider receives a reward based on an incentive model \cite{abusafia2020incentive}.\looseness=-1

 % The microcell energy demand is advertised to providers in terms of requested energy for the whole day, e.g., from [09:00-15:00]  (See Fig.\ref{fig:fullscenario}). The advertised requested energy is posted for each time slot, e.g. each hour, in terms of amount, duration, and rewards. For example, the first time slot in Fig.\ref{fig:fullscenario} is advertised with energy requirement = 400 (mAh), reward = 10 and the time [09:00-10:00]. We assume this advertisement is announced through a  registry. We assume that the announcement is posted at the close of business (COB) time a day ahead of the service day.  We also assume providers register their services a night before the service day. Providers register their services in terms of amount and availability in time slots. Energy providers may offer their services for one or more time slots.  For instance in Fig.\ref{fig:fullscenario} the energy service (ES2) is registered with amount = 600 (mAh) and availability in time slots 1 and 2 from [09:00-11:00]. The IoT coordinator will select and compose energy services to serve as many consumers as possible, aiming to increase their quality of experience in the microcell.

The super-provider will allocate services to time slots to serve as many consumers as possible to \textit{maximize} their quality of experience in the microcell. However, it is challenging to fulfill multiple energy requirements with limited energy services \cite{lakhdari2021fairness}.  For example, in Fig.\ref{fig:fullscenario} (A), the total energy demand ($\sum R$) is 3200 mAh, and  the total available energy services ($\sum S$) is 2300 mAh. The available services may fulfill 71.9\% of the energy requests which cannot be fully provisioned with the temporal constrains of services and requests. Therefore, satisfying all consumers with their under-provisioned requests is\textit{ more challenging.}\looseness=-1 

Fig.\ref{fig:fullscenario} (B) presents the outcome of a greedy FCFS, i.e., first come first served, allocation strategy for the available energy \cite{kruse2007data}.  In greedy, the time slots and services will be scheduled based on their start time. For instance, in Fig.\ref{fig:fullscenario} (A) even though $S2$ can be offered in $t_1$, $t_2$ and $t_3$, $S1$ will be allocated to $t_1$ because it comes first in terms of time. The greedy strategy does not leverage the shareable nature of energy services or the providers' flexibility which may affect the energy allocation efficiency and impact the consumers' experience. Therefore, the greedy strategy may not be a good fit in this context. For example, in Fig.\ref{fig:fullscenario} (B), the greedy-based approach could only fulfill 1300 mAh from the total demand which is equivalent to 40.6\%. Moreover, the total number of consumers ($\sum C$) in Fig.\ref{fig:fullscenario}(A) is 14 and the greedy approach could offer energy to 7 consumers which is equivalent to  50\%. In this context, we consider the size of fulfilled requests and the number of fulfilled consumers in assessing the quality of experience. In this example, using the greedy approach resulted in  45.3\% of consumers' QoE. 
%The partial-based and demand-based allocation strategies are a good illustration of the effect of a QoE-aware provisioning plan. Both strategies achieved  73.8\% and 79.7\% of enhancing the QoE. Additionally, the demand-based approach satisfied 88.2\% of the consumers, which is higher than all other strat

Allocating the limited available energy with the time constraints of both services and requests represents critical challenges for efficient and QoE-aware provisioning of IoT energy services. We propose a framework that will compose the energy services to maximize  the consumers' experience. Our framework leverages leverage the providers' \textit{flexibility} and  \textit{shareable} nature of energy services to efficiently allocate services and optimize the QoE.

\vspace{-15pt}
\section{Preliminaries}
\vspace{-8pt}

 %We represent a formal model of our composition of energy requests. 
  We consider the scenario of energy sharing in a microcell $M$ during a time interval \emph{T}. \emph{T} is divided into a set of $\{ t_1, ..., t_n\}$ where $t_i$ is a predefined time period, e.g., one hour. We use the below definitions to formalize the problem.

\textbf{Definition 1:  Energy-as-a-Service (EaaS).} We adopt the definition of $EaaS$ in \cite{lakhdari2018crowdsourcing}.  An energy service ($EaaS$) is a tuple of $<E_{id}, E_{pid}, F,Q>$, where $E_{id}$ is an energy service ID,    $E_{pid}$ is a provider ID, $F$ is the function of sharing wireless energy, $Q$ is a set of non-functional ($QoS$) attributes, including:
        \begin{itemize}[ noitemsep,nosep,leftmargin=10pt,labelsep=2pt,itemindent=2pt, labelwidth=*]
            \item $p_{ae}$ is the amount of energy shared by the provider.
            \item $p_{loc}$ is the location of the energy provider $<x,y>$.
            % \item $p\_{st}$ is the start time of the EaaS. 
            % \item $p\_{et}$ is the end time of the EaaS. 
            \item $p_t$ is the set of time intervals $<t_s,t_e>$ a provider may offer their energy. 
        \end{itemize}

 \textbf{Definition 2: Energy Service Request (ER).} We adopt the definition of ER in \cite{abusafia2020incentive}.  An  $ER$ is a tuple of $<E_{id}, E_{cid}, F,QR>$, where $ER_id$ is an energy request ID,  $E_{cid}$ is a consumer ID, $F$ is the function of receiving energy wirelessly by an IoT device, $QR$ is a set of non-functional attributes, including:\looseness=-1
    \begin{itemize}[ noitemsep,nosep,leftmargin=10pt,labelsep=2pt,itemindent=2pt, labelwidth=*]
        \item $c_{re}$ is the amount of requested energy.
        \item $c_{loc}$ is the  location of the energy consumer $<x,y>$.
        % \item $c\_{st}$ is the start time of the ER.
        % \item $c\_{et}$ is the end time of the ER.
         \item $c_t$  is the time interval $<t_s,t_e>$ of requiring energy.
\end{itemize}

% \textbf{Definition 3: Incentive.} 
% An energy provider earn a reward $rwd$ by sharing energy.  Rewards $R$ are granted by the IoT coordinator in the form of stored credits.  We assume the system limits the amount of requested energy. The earned credits may be used later by the provider to increase the limit of requested energy as a consumer. A study on the use of incentives to increase energy provision participation has been conducted \cite{abusafia2020incentive}. The study shows that the amount of provided energy impacts the value of the reward. We, therefore, will compute the reward ($rwd$) of an energy request based on the amount of requested energy using the below equation:
% \begin{equation}
%  \label{eq:rwd}
%  rwd  = c_{re} / 100 %* Reward\_Points
% \end{equation}
% where $ c_{re} $ is the amount of requested energy and 100 is the energy capacity of a full charge. 

\textbf{Definition 3: Microcell Energy Demand $\mathcal{MED}$.} 
%The demand of energy in a microcell
$\mathcal{MED}$ is the total amount of requested energy during a time interval $T$  (See Fig.\ref{fig:fullscenario}).  $T$ is divided into time slots. We define $\mathcal{MED}$  by aggregating the amount of required energy per time slot.  Therefore, the definition of $\mathcal{MED} = \{t_1, t_2,...,t_n\}$ where $t$ is a tuple of $< d, rwd, re, nc, ER>$. Here $d$ is a predefined time in the time interval of the microcell $T$, e.g., [9:00 AM -10:00 AM].  $rwd$ is the reward of providing the required energy $re$. We compute $rwd$ using the incentive model proposed by \cite{abusafia2020incentive}. We assume that the super-provider will use the microcell history to compute the energy demand in advance. $nc$ is the number of consumers in the microcell at time slot $t$. $ER$ is the set of available requests in the microcell at time slot $t$.\looseness=-1 

% We compute $re_i$ from the recorded history of microcell using the below equation:
% \begin{equation}
%  \label{pdctREequation}
%  re_i  = \sum_{j =1}^{n} \left( \sum_{k=1}^{m}  c_{re_{jk}} | c_{t_{jk}} \subset t_i  \right)/n
% \end{equation}
%  where $c_{re_{jk}}$ is the amount of requested energy by consumer. $c_{t_{jk}}$ is the time interval of the energy request. $t_i$ is the time slot in the microcell interval $T$. Note that we only consider $c_{re_{jk}}$ which their time interval $c_{t_{jk}}$ fall in the time slot $t_i$.  $m$ is the number of energy requests $ER$ which their time interval fall in the time slot $t_i$.  $n$ is the number of recorded days , e.g, last 90 days.
%  \textbf{Definition 4: Microcell ($M$).} $M$ is a tuple of $<\mathcal{L},\mathcal{MED},T>$, where  $\mathcal{L}$ is the microcell spatial location,  $\mathcal{MED}$ is the microcell energy demand and $T$ is a predefined time interval. 
 
 \textbf{Definition 4: Quality of Experience (QoE).}  $QoE$ is defined as an objective function to measure consumers' satisfaction with energy provisioning in a microcell $M$ within a predefined time interval $T$. 
The function definition is:
\begin{equation} \label{eq:MQoE}
 QoE(M)  =  F(T,\mathcal{ES},\mathcal{MED})
\end{equation} 
where $\mathcal{ES}$ is the set of energy services and  $\mathcal{MED}$ is the microcell energy demand. 
\vspace{-25pt}
\subsection{Problem Definition}
\vspace{-5pt}
Given a set of $n$ energy services  $\mathcal{ES} = \{EaaS_{1},EaaS_{2},....,$ $EaaS_{n}\}$ and a set of $m$ energy requests   $\mathcal{ER} = \{ER_{1},ER_{2},....,$ $ER_{m}\}$ in a microcell $M$. %$\mathcal{ES}$ are submitted by providers $P$ and $\mathcal{ER}$ are initiated by consumers $C$.
The super-provider advertise the microcell energy demand $\mathcal{MED}$. Energy providers register their services in terms of: (1) the amount of energy $p_{ae}$ (2) the time slots $t_i$ to offer their services. The super-provider uses the providers preferences to allocate their services to time slots. The allocation approach aims at fulfilling the maximum number of requests and thereby maximize the QoE. 
%We identify this problem as a resource allocation problem as energy services need to be allocated to time slots.
%We assume providers may offer their service in multiple time slots if they have enough energy. In addition, providers may offer partial services to multiple consumers. We also assume that energy requests $\mathcal{ER}$ are already known from the computed energy demand $\mathcal{MED}$ of a microcell.
We formulate the service composition problem to a time-constrained optimization problem as follows:%, service composition problem. % The composition of energy services considers the features of services and requests and the time-slot preferences of providers.
% The aim is to \textit{maximize the quality of experience in the microcell QoE(M)}. We formally define the spatio-temporal composition as follows:\looseness=-1

%by maximizing: (1) the number of fulfilled requests and (2) the utilization of the provided energy.  We formally define the spatio-temporal composition as an optimization as follows:
\begin{itemize}[ noitemsep,nosep]%leftmargin=10pt,labelsep=2pt,itemindent=2pt, labelwidth=*]
    \item Maximize $QoE(M) =  F(T,\mathcal{ES},\mathcal{MED})$,
  \end{itemize}
 {Subject to:
\begin{itemize}[ noitemsep,nosep]%,leftmargin=10pt,labelsep=2pt,itemindent=2pt, labelwidth=*]
     \item $  t_i.re > 0 $  for each $t_i  \in T$,
    \item $ EaaS_{j}.P_t \subset t_i.d$  for each $EaaS_{j}  \in \mathcal{ES}$.
\end{itemize}
Where    $ P_t$ is the time interval $<t_s,t_e>$ a provider of $EaaS_{j}$  may offer their energy,     $ t_i.d$ is the duration of a time slot $i$ in the time interval of the microcell $T$, and\looseness=-1
   $ t_i.re$ is the required energy $re$ at time slot $i$.
}

\noindent The goal of the composition is to efficiently allocate the available energy services to  time slots. The objective function attempts to optimally assign energy services according to their spatio-temporal features, providers' preferences and required energy in time slots. {The spatial aspect in energy service focuses on a geographical cell. The temporal aspect focuses on the times of energy service provisioning.}\looseness=-1

% For instance, we assume we know the consumers’ energy needs based on previous history [3]. With regard to energy providers, we employ an incentive model which allows us to predict the amount of energy that would be available for consumption [4].
\noindent We use the following assumptions to formulate the problem.
\begin{itemize}[ noitemsep,nosep,leftmargin=10pt,labelsep=1pt,itemindent=0pt, labelwidth=*]
\item Providers energy size is fixed during  composition.
\item Providers are available in all their selected time slots. % and they may offer their service in multiple time slots if they have enough energy.
\item Providers may offer partial services to multiple consumers at the same time.
\item Consumers' time windows do not overlap with time slots.
\item Providers and consumers have fixed location during energy sharing.
%\item A provider transfers energy to consumers in the microcell.
%\item A provider transfers energy to one or more consumers at the same time %using ``Mi Air" technology \cite{mi_blog}. %\footnote{https://www.mi.com/}.
\item The microcell has \textit{multiple} providers and \textit{multiple} consumers.
\item There is no energy loss in sharing. {As the technology matures, we anticipate that the devices will be able to share more energy, and the energy loss of sharing will become minimal \cite{lakhdari2021fairness}}. 
\item The exact amount of required energy for a microcell is given \cite{huang2012predicting}. 
\item A reward system is used to incentivize providers to offer their service \cite{abusafia2020incentive}.
\item A trust framework is used to preserve the  privacy of the IoT devices \cite{lakhdari2020Elastic}.
\end{itemize}

\vspace{-16pt}
\section{Quality of Experience Model}\label{QoETest}
\vspace{-8pt}
The Quality of Experience (QoE) in a microcell is measured based on the \textit{number of satisfied consumers }and the\textit{ amount of fulfilled requests}. Recall, the time interval of the microcell is divided into time slots.  Therefore, QoE for each time slot $t_i$  will be computed using the following attributes:

\begin{itemize}[ noitemsep,nosep,leftmargin=5pt,labelsep=1pt,itemindent=0pt, labelwidth=*]
\item \textbf{Satisfaction Ratio:} 
 We define the Satisfaction Ratio ($\mathcal{SR}$) as the number of consumers who received their requested energy or part of it.  We compute $\mathcal{SR}$ per time slot $t$ as follows:\looseness=-1
 
 \begin{equation}
 \label{eq:ss}
    \mathcal{SR} = \frac{|\{ER \in \mathcal{ER} \;|\; ER\;is\; completed\; \&\; c_t \in d\}|}{|\mathcal{ER}|}
\end{equation}
 Where $\mathcal{ER}$ is the set of all requests in time slot $t$, $|.|$ is the cardinality of the set, $c_t$ is the request time, and $d$ is the time duration of $t$. %The value of $\mathcal{SR}$ increases when the number of served requests  approaches the total set of energy requests. 
\item \textbf{Fulfillment Ratio:} 
The satisfaction ratio is not enough to measure QoE. For example, if we have a set of energy requests in mAh $\mathcal{ER}$= \{10, 20, 20, 70\}, serving the first 3 consumers is not equal to serving the last 3 due to the different amount of requested energy. Therefore, We define the Fulfillment Ratio ($\mathcal{FR}$) based on the percentage of fulfillment for each request. We compute $FR$ per time slot $t$ as follows:
%For instance, a consumer who got 80\% of their request will be more satisfied than a consumer who got 30\% of their request. We compute the $SL$ per time slot $t_i$ as follows:
  %\mathcal{SL}_i=   \left(\sum_{i=1}^{n} Received\_Energy_{i} \right)  /  de_i.re
\begin{equation}
\label{eq:sl}
    \mathcal{FR}=   \sum_{i=1}^{n} \left(w_i \times \frac{ Received\_Energy_{i}}{Requested\_Energy_{i}} \right) 
\end{equation}
where \textit{n} is the number of all energy requests in $t$, and $w_i$ is the weight of the request over the total amount of requested energy in $t$. % The previous equation measures the ratio of the fulfillment of all consumers in time slot $t$.

\end{itemize}
 
% \noindent\textbf{ Quality of Experience (QoE).}  As previously stated, the quality of experience of a time slot $t_i$ will be computed based on the satisfaction size $\mathcal{SS}$ and Satisfaction Level $\mathcal{SL}$. 

% Therefore, we compute the quality of experience of a time slot ($QoE(t_i)$) as the following:
% \begin{equation}
% \label{eq:tsQoE}
% QoE(t_i) =  \mathcal{SS}_i +\mathcal{SL}_i
% \end{equation}
% where $\mathcal{SS}_i$ is the satisfaction size of a time slot $t_i$ computed by Eq.\ref{eq:ss} and $\mathcal{SL}_i$ is the satisfaction level of time slot $t_i$ computed by Eq.\ref{eq:sl}.

\noindent\textbf{Quality of Experience:} 
As previously stated, We define the QoE in a microcell based on the satisfaction ratio $\mathcal{SR}$ and fulfillment ratio $\mathcal{FR}$ of each time slot $t$. Therefore, we compute the $QoE(M)$ as the following:

\begin{equation}
%\resizebox{1.1\hsize}{!}{$
\label{eq:fullQoE}
QoE(M) =  \alpha \times \left(\sum_{i=1}^{m} SR_i \times \beta_i \right) + (1-\alpha) \times \left(\sum_{i=1}^{m} FR_i\times \gamma_i \right) %$}
\end{equation}
where \textit{m} is the number of time slots in the microcell's time interval $T$. $\mathcal{SR}_i$ is the satisfaction ratio of a time slot computed by Eq.\ref{eq:ss}. $\beta_i$ is the weight of a time slot $t_i$ which is its number of  consumers over the total number of consumers in $T$. $\mathcal{FR}_i$ is the  ratio of fulfillment of the time slot computed by Eq.\ref{eq:sl}. $\gamma_i$ is the weight of a time slot $t_i$ which is its total required energy over the total amount of required energy in  $T$. $\alpha$ is a user-defined weight between zero and one to define the weight of  $\mathcal{SR}_i$  and $\mathcal{FR}_i$ in $QoE$.\looseness=-1
% based on the preference of the microcell owner.
\begin{figure}[!t]
    \centering
       \setlength{\abovecaptionskip}{5pt}
     \setlength{\belowcaptionskip}{-15pt}
    \includegraphics[width=0.65\linewidth]{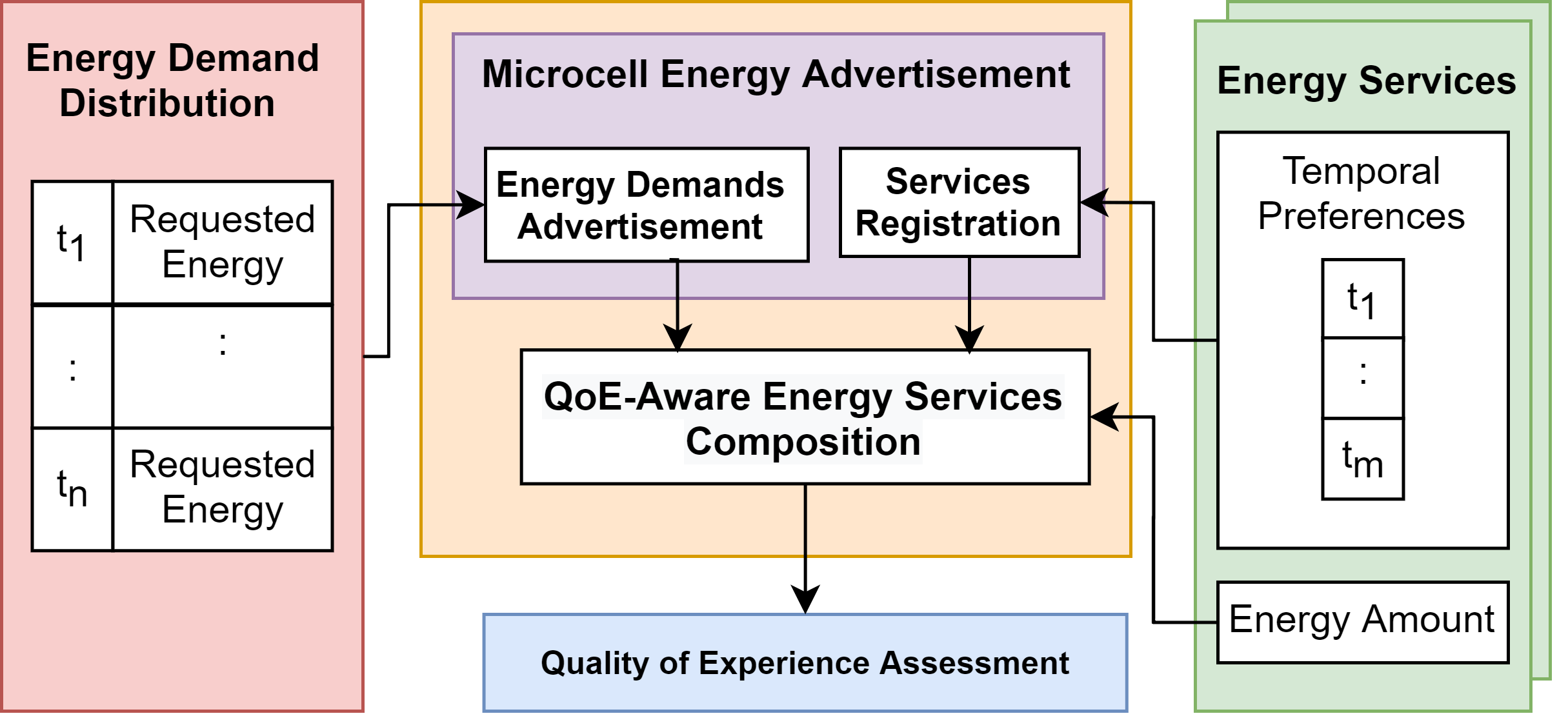}
    \caption{Quality of experience driven service composition framework }
    \label{fig:frameWork}
  \end{figure}
  \vspace{-15pt}
\section{Quality of Experience Framework}
  \vspace{-10pt}
We introduce a quality of experience composition framework for managing energy services to enhance consumers’ QoE (See Fig.\ref{fig:frameWork}). The framework is divided into three phases: (1) Microcell energy advertisement, (2) Composing energy services, and (3) Quality of experience assessment. In the first phase, the super-provider will advertise the energy demand of the microcell and receives providers’ preferences. In the second phase, the super-provider will compose energy services to maximize the QoE. In the last phase, the super-provider will assess the QoE for the resulted composition.\looseness=-1
\vspace{-15pt}
\subsection{Microcell Energy Demand Advertisement}
\vspace{-8pt}
In this phase, the super-provider computes the reward for each time slot based on the amount of required energy using the incentive model in \cite{abusafia2020incentive}. Then, the system will announce the required energy and rewards for the whole microcell using Definition 3. Energy providers will register based on their preferences in terms of their energy amount and the time slots they will be available (See Fig.\ref{fig:fullscenario}).
\vspace{-25pt}
\subsection{Energy Services Composition}
\vspace{-8pt}
This phase aims to compose energy services to maximize the QoE. We propose two heuristic approaches to compose energy services: \textit{Partial-Based} (PB) and \textit{Demand-Based} (DB). The PB composition is inspired by the FCFS resource allocation algorithm \cite{kruse2007data}. The PB approach, splits services among consumers if the required energy is more than the available energy. Intuitively, offering part of the services will satisfy more consumers than offering it to some of them. The DB composition is inspired by the priority allocation algorithm \cite{kruse2007data}. The DB approach  extends the PB approach by prioritizing slots with the highest demanding to ensure services availability. We discuss each approach below.\looseness=-1

\begin{algorithm}[!t]
 \caption{ Partial-Based Composition of Services}
 \label{alg:PB}
 \footnotesize
 \begin{algorithmic}[1]
 \renewcommand{\algorithmicrequire}{\textbf{Input:}}
 \renewcommand{\algorithmicensure}{\textbf{Output:}}
 \REQUIRE $\mathcal{MED}$, $\mathcal{ES},threshold$
 \ENSURE  $energy\_comp$% , compRwd$
%  \STATE ${Rwds}$ \textbf{= Compute\_Rewards($\mathcal{MED}$)} \STATE $Bidding$ \textbf{= Energy\_Demand\_Advertisement($\mathcal{MED},{Rwds}$)}
 %\STATE $QoE$ = 0
%  \STATE $totalEnergy$ = 0
 \FOR {$t_{i}$ in $\mathcal{MED}$} 
    \STATE  $selectedES =\{\}$ \STATE $demand$ = $t_i.re$
    \FOR {$es_{j}$ in $t_i.\mathcal{ES}$} 
    \IF{$demand >$ 0}
        \STATE $demand$ = $demand$ - $es_j.p_{ae}$
        %\STATE $totalEnergy$ = $totalEnergy$ + $es_j.p_{ae} $
        \STATE $energy\_comp.add(t_{i},es_j)$
         \STATE $selectedES.add(es_j)$
        \IF{$demand <$ 0}
            \STATE $es_j.p_{ae}$ =$demand$ * -1 
            \STATE $demand = 0$
        \ELSE
        \STATE \textbf{Remove\_Service($es_{j},\mathcal{MED}$)}
        \ENDIF 
    % \ELSIF{$demand$ = 0}
    
    % \STATE \textbf{break}
    \ENDIF
    \ENDFOR
    \IF{$demand $ = 0}
            \STATE \textbf{Assign\_Energy($t_i.\mathcal{ER},selectedES$)}
        \ELSE
        % \STATE $nc$ = $t_i.{nc}$
        % \STATE $Energy$ = $totalEnergy$ / $nc$  
        % \WHILE {$Energy < threshold$}
        %     \STATE $nc$ = $nc$ - 1
        %      \STATE $Energy$ = $totalEnergy$ / $nc$  
        % \ENDWHILE
         \STATE \textbf{Assign\_Partial\_Energy($t_i.\mathcal{ER}, selectedES , nc, threshold$)}
        \ENDIF
 \ENDFOR
% \STATE $compQoE$  \textbf{= Compute\_Quality\_Experience($energy\_comp$)}
  % \STATE $compQoE$ = \textbf{Compute\_QoE($ \mathcal{T},\mathcal{EDD},energy\_comp$)}
 \STATE \textbf{return }$energy\_comp $
 \end{algorithmic}
 \end{algorithm}
 \setlength{\textfloatsep}{4pt}

\vspace{-15 pt}
\subsubsection{Partial-Based Energy Services Composition} \label{PBsection}

The Partial-Based (PB) composition aims at maximizing the QoE by composing services for each time slot based on the first come first served approach. For example, if a provider offers their services on two-time slots, the algorithm will assign the service for the earlier time slot. If the time slot did not need the service, the service will be assigned to the next time slot. Moreover, PB chunks services between energy consumers if the available energy services are not enough to fulfill the total required energy in the time slot. Intuitively, offering part of the required services to all consumers is more satisfying than offering it only to some of them.\looseness=-1

Algorithm \ref{alg:PB} presents the PB service composition. For every time slot $t_i$, the algorithm retrieves the total required amount of energy (Line 3). Then, for each registered service $es$ in $t$, the algorithm keeps adding services to the set of selected services until the required energy is fulfilled or all the available services have been selected (Lines 4 - 13). Note that if a service was partially needed, then the service available amount will be updated to be used by other registered time slots (Lines 9 - 11). Moreover, if a service was fully used by a time slot, then it will be removed from other registered time slots (Lines 12 - 13). After processing all services, if the energy demand of the slot is zero, the algorithm assigns the selected services to requests (Lines 14 - 15). If the energy demand is not fulfilled,  the algorithm distributes the available services among available requests (Line 17). % Given that the service chunks are greater than a predefined threshold. The threshold prevents dividing services into small neglectable chunks.
If the service chunks are smaller than the threshold, consumers will be removed and the service will be shared among the rest. The threshold prevents dividing services into small neglectable chunks.
% This step will be repeated until the chunks are greater than or equal to the threshold (Lines 21 - 25). Then, the services will be assigned to the available energy requests based on the size of the chunks (Line 26).
%Once the allocation of the service is completed, a quality of experience function will be invoked to asses the resulted composition (Line 18).
The composition  of the selected services will be returned in Line 18. \vspace{-15pt}
\subsubsection{Demand-Based Energy Services Composition}
The Demand-Based (DB) composition goal is to maximize QoE by giving priority to time slots with higher energy demand. The intuitive idea of the DB approach is that high-demanding time slots will require more services. Thus, services should be assigned to them prior to less demanding time slots which may ensure a better distribution of the available services.  For instance, if a provider offers their service on two-time slots, the algorithm will assign the service to the more demanding time slot. If that time slot does not need the service, the service will be assigned to the next time slot. This indicates that the order of time slots in composing services matters because if a service is used in a time slot, it will be removed from others. Removing a service from a time slot may affect the amount of available energy and thus the number of served and satisfied consumers. Moreover, DB approach maximizes the QoE by chunking services between energy consumers if the available services are not enough.\looseness=-1

\begin{algorithm}[!t]
 \caption{Demand-Based Composition of Services}
 \label{alg:DB}
\footnotesize
 \begin{algorithmic}[1]
 \renewcommand{\algorithmicrequire}{\textbf{Input:}}
 \renewcommand{\algorithmicensure}{\textbf{Output:}}
 \REQUIRE $\mathcal{MED}$, $\mathcal{ES},threshold$
 \ENSURE  $energy\_comp$% , QoE $ , compRwd$
%  \STATE ${Rwds}$ \textbf{= Compute\_Rewards($\mathcal{MED}$)} \STATE $Bidding$ \textbf{= Energy\_Demand\_Advertisement($\mathcal{MED},{Rwds}$)}
  %\STATE $totalEnergy$ = 0
   \STATE $\mathcal{SMED}$ = \textbf{sort($\mathcal{MED},nc:descending,re:descending$)}
 \FOR {$t_{i}$ in $\mathcal{SMED}$} 
    \STATE  $selectedES =\{\}$
    \STATE $demand$ = $t_i.re$;
    \STATE $sortedES$ = \textbf{sort($t_i.\mathcal{ES},nt:ascending$)} 
    \FOR {$es_{j}$ in $sortedES$} 
    \IF{$demand >$ 0}
        \STATE $demand$ = $demand$ - $es_j.p_{ae}$
        %\STATE $totalEnergy$ = $totalEnergy$ + $es_j.p_{ae}$
        \STATE $energy\_comp.add(t_{i},es_j)$
        \STATE $selectedES.add(es_j)$
        \IF{$demand <$ 0}
            \STATE $es_j.p_{ae}$ =$demand$ * -1 
            \STATE $demand = 0$
        \ELSE        \STATE \textbf{Remove\_Service($es_{j},\mathcal{SMED}$)}
        \ENDIF
    % \ELSIF{$demand$ = 0}
    %     \STATE \textbf{break}
    \ENDIF
    \ENDFOR
     \IF{$demand $ = 0}
            \STATE\textbf{Assign\_Energy($t_i.\mathcal{ER},selectedES$)}
        \ELSE
        % \STATE $nc$ = $t_i.{nc}$
        % \STATE $Energy$ = $totalEnergy$ / $nc$  
        % \WHILE {$Energy < threshold$}
        %     \STATE $nc$ = $nc$ - 1
        %      \STATE $Energy$ = $totalEnergy$ / $nc$  
        % \ENDWHILE
        %\STATE $RemainEnergy$=\textbf{Assign\_Partial\_Energy($t_i.\mathcal{ER},$        \item[] \hspace{3cm} $selectedES , nc$)}
        \STATE \textbf{Assign\_Partial\_Energy($t_i.\mathcal{ER},selectedES , nc, threshold$)}
        % \IF{$RemainEnergy > $ 0}
        %     \STATE $nc$  \textbf{  = Get\_Consumers($\mathcal{MED}$)}
        %      \STATE \textbf{Go to Line $26$ }
        % \ENDIF
    \ENDIF
 \ENDFOR
 %\STATE $compQoE$  \textbf{= Compute\_Quality\_Experience($energy\_comp$)}
 
 \STATE \textbf{return }$energy\_comp$
 \end{algorithmic}
 \end{algorithm}

Algorithm \ref{alg:DB} presents the DB service composition.  The algorithm starts by sorting the time slots in a descending order based on the number of consumers $nc$, then the  amount of requested energy $re$ (Line 1). The goal of sorting is to start composing services for the most demanding time slots. As some services may be registered in multiple services, using these services for the most demanding time slots may offer a better experience. Line 4 retrieves the total required amount of energy for each time slot $t_i$. Then, for every time slot, the registered services will be sorted in ascending order based on the number of time slots a service was registered in. This sort will allow us to start with the least connected services. In other words, using such services may impact less number of time slots than using services that are registered in many time slots. Then,  for each registered service $es$ in $t$, the algorithm keeps adding services to the set of selected services until the required energy is fulfilled or all the available services have been selected (Lines 6 - 15). Similar to the PB approach, if a service was partially needed, then the service available amount will be updated to be used by other registered time slots (Lines 11 - 13). Moreover, if a service was fully used by a time slot, then it will be removed from other registered time slots (Lines 14 - 15). After processing all services, if the energy
 demand of the slot is zero, the algorithm assigns the selected services to requests (Lines 16 - 17). If the energy demand is not fulfilled,  the algorithm distributes the available services among available requests (Line 19). % Given that the service chunks are greater than a predefined threshold. The threshold prevents dividing services into small neglectable chunks.
If the service chunks are smaller than the threshold, consumers will be removed and the service will be shared among the rest. The threshold prevents dividing services into small neglectable chunks.
Line 20 returns the composition of the selected services.\looseness=-1
% After going through all energy services, the algorithm checks if the energy demand is zero, then the services will be assigned to the requests (lines 20 - 21). If the demand of the time slot is not fulfilled,  the algorithm shares the available services among available requests equally if the chunks are greater than a predefined threshold (line 28). The threshold prevents dividing services into small neglectable chunks. If the chunks were smaller than the threshold, a consumer will be removed and the service will be shared among the rest. This step will be repeated until the chunks are greater than or equal to the threshold (Lines 23 - 27). Then, the services will be assigned to the available energy requests based on the size of the chunks. If a request requires less energy than the assigned chunk, the algorithm will aggregate the remaining energy and redistribute it among consumers who are still in need of energy (Lines 28 - 31).
 
%\setlength{\textfloatsep}{0pt}

\vspace{-15 pt}
 \subsection{ Assessing Quality of Experience}
 \vspace{-8 pt}
The super-provider assesses the QoE of each proposed composition in this phase. The QoE is computed using the model discussed in Section \ref{QoETest}. The assessment of QoE gives an indicator of consumers' satisfaction in the microcell.  
\vspace{-15pt}
\section{Evaluation}
\vspace{-10pt}

%Our proposed framework is the first attempt to allocate energy services to increase the QoE. We reformulate our service provisioning problem as a time-constrained resource allocation problem, i.e., maximizing the QoE by efficiently allocating the available energy.
We compare the proposed composition approaches, Partial-Based composition (PB), and  Demand-Based Composition (DB), {with the resource allocation algorithms, namely, first come first served allocation (\textit{Greedy}), and  Max-Min Fair allocation (\textit{Max-Min}) \cite{kruse2007data}\cite{lakhdari2021fairness}. The \textit{Greedy} approach is a modified FCFS algorithm where the time slots and services will be scheduled based on their start time. The \textit{Max-Min} is a modified Max-Min Fair allocation where services that can be offered in multiple time slots will be split among these time slots using the a Max-Min technique. We evaluate the \textit{effectiveness} and the \textit{efficiency} of each approach.}\looseness=-1

\vspace{-15pt}
\subsection{Dataset Description}
\vspace{-8pt}

We used a real dataset generated from the developed app in \cite{yao2022wireless}. The dataset consists of energy transfer records between a provider (smartphone) and a consumer (smartphone). The records attributes are the provider ID, consumer ID, transaction date, time, energy services' and requests' amount, and transfer duration. We use the energy dataset to generate the QoS parameters for the energy services and requests. For instance, the amount of a wireless charging transfer in mAh is used to define the amount of requested/provided energy. In addition,  the energy dataset records of a wireless charging transfer duration are used to define the end time of each request/service. \looseness=-1 

We augmented the dataset of the energy sharing to mimic the behavior of the crowd within microcells by utilizing a dataset published by IBM for a coffee shop chain with three branches in New York city\footnote{https://ibm.co/2O7IvxJ}. The dataset consists of transaction records of customers purchases in each coffee shop for one month. Each coffee shop consists of, on average, 560 transnational records per day and 16,500 transaction record in total. We use the IBM dataset to simulate the spatio-temporal features of energy services and requests. Our experiment uses the consumer ID, transaction date, time, location, and coffee shop ID from each record in the dataset to define the spatio-temporal features of energy services and requests, e.g., start and location of energy service or a request.  We ran a total of 7000 experiments with 6-time slots each time slot was an hour long. In each run, the providers' temporal provision preferences were registered randomly to [1-3] time slots. In addition, the number of services and requests varied between 300 to 2000 per run depending on the experiments' setting. For each run, we used the proposed approaches to compose energy services.  We then measured the QoE for each composition.  Table \ref{ExpVariables} presents the experiments parameters.\looseness=-1

\vspace{-15pt}
\subsection{Evaluation of the Composition Framework}
\vspace{-10pt}
We ran six experiments to determine the effectiveness and efficiency of the proposed approaches. The experiments evaluated the approaches in terms of their satisfaction rate, fulfillment rate, quality of experience, impact of thresholds and computation cost. We run the approaches in different settings by changing the ratio of services to requests in the time interval $T$. We gradually increased the ratio from 15\% to 90\%. We repeated the experiment 1000 times at each point and considered the average value for each approach. % at each point.
% where the ratio of services to requests increased.

\begin{table}[!t]
\noindent
    \begin{minipage}{\linewidth}
      \begin{minipage}[b]{0.5\linewidth}
        \centering
        
        \caption{Experiments Variables}
        \label{ExpVariables}
    %       \setlength{\abovecaptionskip}{10pt}
    % \setlength{\belowcaptionskip}{-12pt} 
    % \renewcommand{\arraystretch}{1}
    % \scalebox{0.75}{
        \resizebox{\linewidth}{!}{%
        \renewcommand{\arraystretch}{1}
        \begin{tabular}{|l|l|}
        \hline
        Variables & Value   \\ \hline
        Energy dataset for coffee shop 8 in April            & 16830   \\ 
        Number of services \& requests           & [300-2000]/run \\ 
        Number of time slots       & {6}   \\
        Provided energy                  & {5 - 100\%}      \\
        Requested energy                 & {5 - 100\%}       \\ 
        Time interval             & {6 hours} \\ 
        Service registration & [1-3] time slots / service\\\hline
        \end{tabular}}%
          \end{minipage}
      \hfill
     \begin{minipage}{.5\linewidth}
        \centering
       
         \setlength{\abovecaptionskip}{2pt}
             \setlength{\belowcaptionskip}{-5pt}
        \includegraphics[width=0.8\linewidth]{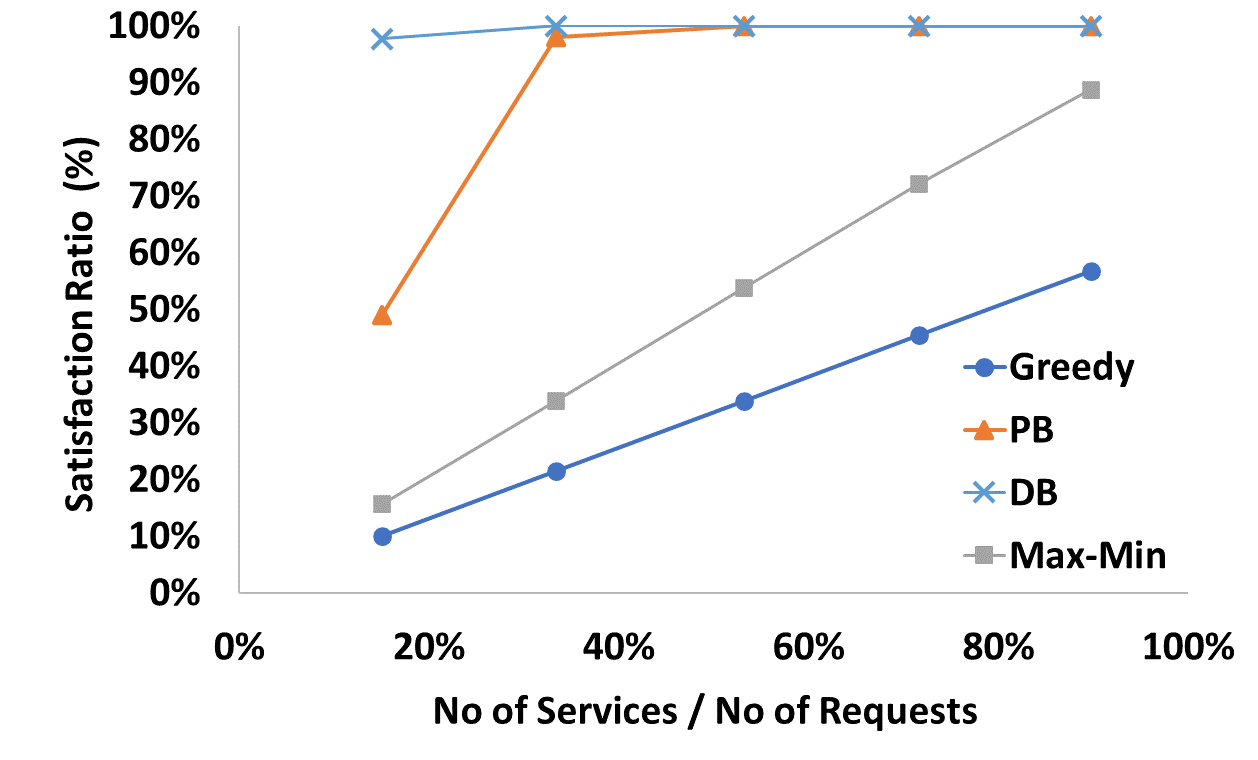}
         \captionof{figure}{The average of satisfaction ratio}%in a microcell}
         \label{fig:expss}
        \end{minipage}
            \end{minipage}
\end{table}

\vspace{-15 pt}
\subsubsection{Quality of Experience Evaluation}
\begin{figure}[!t]
\centering
\begin{minipage}{.48\textwidth}
\centering
 \setlength{\abovecaptionskip}{0pt}
     \setlength{\belowcaptionskip}{-5pt}
\includegraphics[width=1\linewidth]{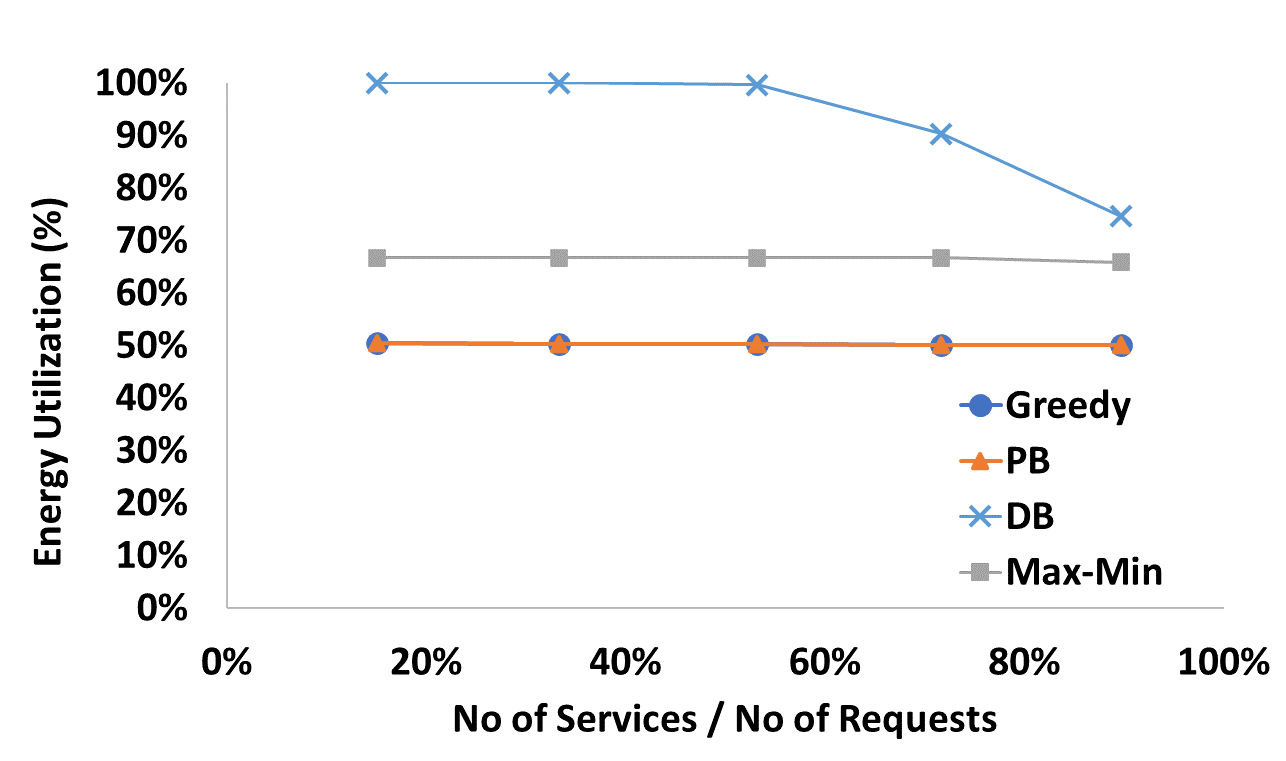}
\caption{The average of energy utilization }
\label{fig:expEU}
\end{minipage}
\hfill
\begin{minipage}{.48\textwidth}
\centering
 \setlength{\abovecaptionskip}{0pt}
     \setlength{\belowcaptionskip}{-5pt}
\includegraphics[width=1\linewidth]{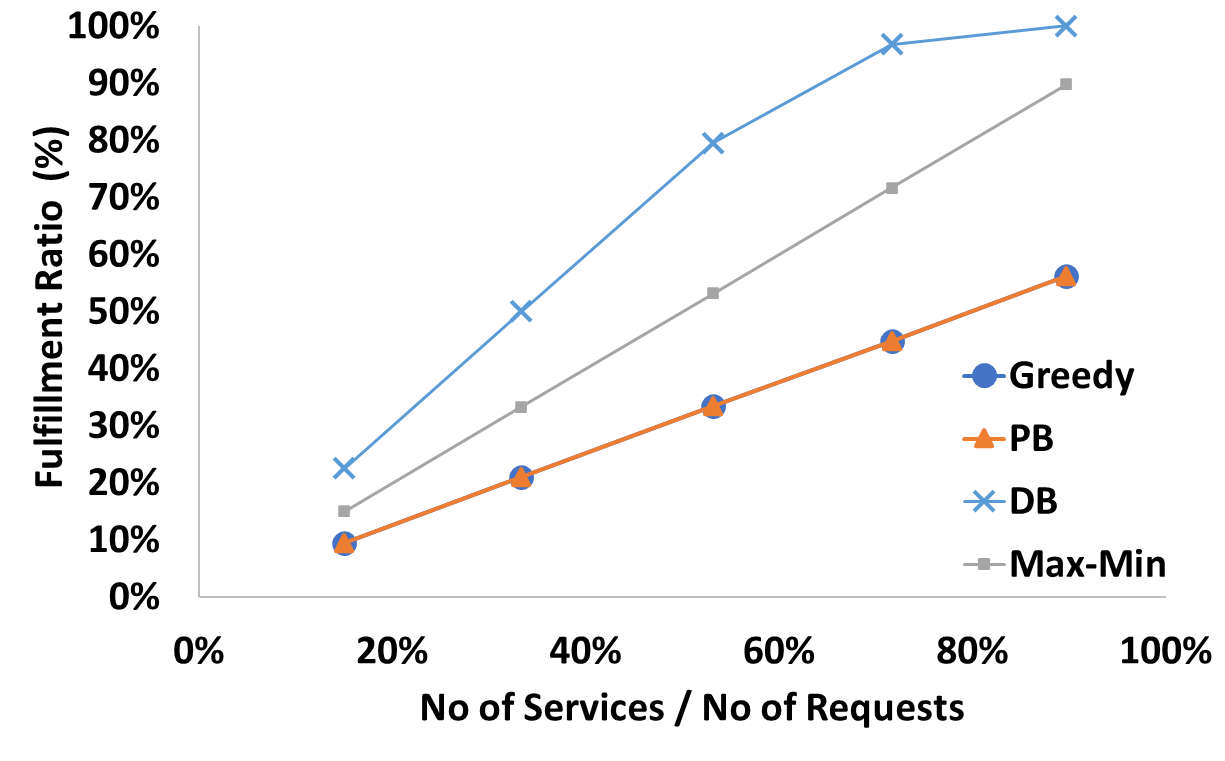}
\caption{The average of fulfillment ratio}%  in a microcell}
\label{fig:expsl}
\end{minipage}      

%\qquad

\end{figure}

As previously stated, we compute the QoE based on $\mathcal{SR}$ and $\mathcal{FR}$ (See Sec.\ref{QoETest}). In this subsection, we study the impact of each ratio, then we evaluate the QoE. %using Eq.\ref{eq:fullQoE}. %., to evaluate the overall QoE in a microcell. %as a combination of both
\\ \indent The first experiment compares the $\mathcal{SR}$ of the proposed approaches PB and DB, against Greedy and Max-Min. As previously stated, $\mathcal{SR}$ represents the number of consumers who received energy fully or partially. Therefore, a high $\mathcal{SR}$ of a composition ensures a higher number of satisfied consumers and thereby a better QoE. The $\mathcal{SR}$  of a time slot is computed using  Eq.\ref{eq:ss} and then averaged for the microcell similar to the first part of Eq.\ref{eq:fullQoE}. Fig.\ref{fig:expss} presents the average $\mathcal{SR}$ in the microcell for each approach. The x-axis in Fig.[\ref{fig:expss}-\ref{fig:expMQoE}] {represents the ratio of the number of energy services to requests}.  In Fig.\ref{fig:expss}, the $\mathcal{SR}$ increases when the number of available services increases for all the composition approaches. For instance, when the ratio of services to requests is 80\%, all approaches provide a higher $\mathcal{SR}$ compared to the ratio is 20\%. This observation can be explained by the availability of services to offer energy. The more services available, the more requests can be fulfilled. The proposed approach PB performs better than Greedy % in terms of $\mathcal{SR}$ 
as it splits the available energy between the consumers as partial services, unlike the Greedy approach which fulfills a request fully before serving the next request. {For the same reason PB also performs better than  Max-Min. Even though, Max-Min  has a better energy utilization by splitting energy services fairly between time slots (See Fig.\ref{fig:expEU}), a fair distribution of energy does not necessarily result in equally satisfied consumers as in the time slots. This is due to the different energy requirements of consumers}. In addition, the proposed approach DB gives the best results as it prioritizes the time slots that have the highest demand in terms of the number of consumers and amount of required energy. Recall the order of time slots in composing services is crucial because if a service is used in a time slot, it will be removed from others. Removing a service from a time slot may affect the amount of available energy and thus the number of served consumers. Prioritizing the most demanding time slots allows  DB  to have more services to use, and therefore increases $\mathcal{SR}$ by increasing the number of fulfilled consumers.\looseness=-1

%In addition, the BF approach gives slightly better results than RB as it looks for all the possible combinations of compositions, and it selects the composition with the highest reliability score. However, the BF approach has a higher computation cost compared to RB as shown in Fig.\ref{ExecutionTime}.
\indent The second experiment compares the $\mathcal{FR}$ of each approach. %es PB and DB, against Greedy. 
As previously stated, $\mathcal{FR}$ presents the rate of fulfillment for each request. Therefore, a high $\mathcal{FR}$ of a composition ensures a higher level of satisfaction for consumers and thereby a better QoE. The $\mathcal{FR}$ of a time slot is computed using  Eq.\ref{eq:sl} and then averaged for the microcell similar to the second part of Eq.\ref{eq:fullQoE}. Fig.\ref{fig:expsl} represents the average $\mathcal{FR}$ in the microcell for each approach. In Fig.\ref{fig:expsl}, the $\mathcal{FR}$ increases when the number of available services increases for all the approaches. %For instance, when the ratio of services to requests is 80\%, all approaches provide a higher $\mathcal{FR}$ compared to the ratio of 20\%. 
This observation can be explained by the availability of services to offer energy. %The more services available, the more requests can be fulfilled. %Therefore, the $\mathcal{FR}$ of the microcell will increase. 
{PB performs similar to Greedy in terms of $\mathcal{FR}$. This is an expected behaviour since both approaches start with the same time slots and, therefore, have the same set of available services. The difference between both approaches is in the way they share energy among consumers, i.e., complete services in Greedy and partial services in PB. Moreover, Max-Min  has a better $\mathcal{FR}$ because it has better energy utilization (see Fig.\ref{fig:expEU}). A higher energy utilization is achieved  by splitting energy services fairly between time slots.} DB gives the best results as it prioritizes the time slots that have the highest demand as discussed in the previous experiment.\looseness=-1
% in terms of the number of consumers and the amount of required energy.Recall that the order of time slots in composing services matters because if a service is used in a time slot, it will be removed from others. Removing a service from a time slot may affect the amount of available energy and thus the number of served and satisfied consumers. Prioritizing the most demanding time slots allows having more services to use, and therefore increases $\mathcal{FR}$ by increasing the amount of fulfilled energy.

The third experiment compares the QoE using all approaches. % PB and DB, against Greedy. 
As previously stated, the QoE presents the overall satisfaction of consumers across time. Therefore, a high QoE of a composition indicates a higher level of satisfaction for consumers. The QoE is computed using  Eq.\ref{eq:fullQoE}. Note that we used $\alpha = 0.5$ to give equal weight for both $\mathcal{SR}$ and $\mathcal{FR}$. Fig.\ref{fig:expMQoE} presents the average QoE using each approach. In Fig.\ref{fig:expMQoE}, similar to the previous experiments, the QoE increases when the availability of services increase. %the number of available services increases for all approaches. %This is due to the availability of services to offer energy.
%The more services available, the more requests can be fulfilled. Thus, the QoE will increase. 
PB approach performs better than Greedy in terms of QoE due to its higher $\mathcal{SR}$ as discussed in the first experiment. {Additionally, PB preforms better than Max-Min when the number of energy services is less than the request. This is because in a limited resources environment PB will satisfy more consumers (higher $\mathcal{SR}$) by partially fulfilling their requests. However, when there is enough services, Max-Min will better utilize the energy to completely fulfill requests (higher $\mathcal{FR}$).} Moreover, the DB approach gives the best results due to its higher $\mathcal{SR}$ and $\mathcal{FR}$.% Similar to the previous experiments, DB prioritizes the time slots with the highest demand 
%in terms of the number of consumers and the amount of required energy. Prioritizing the highest demanding time slots which allows having more services to use in these slots.\looseness=-1 %, and as a result, increases the QoE.

\vspace{-15 pt}

\vspace{-10 pt}
\subsubsection{Threshold Impact Evaluation}
The following two experiments study the impact of thresholds on the PB approach. Recall that  PB and DB approaches split energy between consumers based on a defined threshold. The threshold prevents dividing services into small neglectable chunks. The experiments of both PB and DB gave the same behavior. Thus,  we are only presenting the results of PB. % In the following experiments, we vary the threshold of partial services from 10\% to 90\%.
 
\begin{figure}[!t]
\centering
\begin{minipage}{.48\textwidth}
\centering
 \setlength{\abovecaptionskip}{0pt}
     \setlength{\belowcaptionskip}{0pt}
\includegraphics[width=1\linewidth]{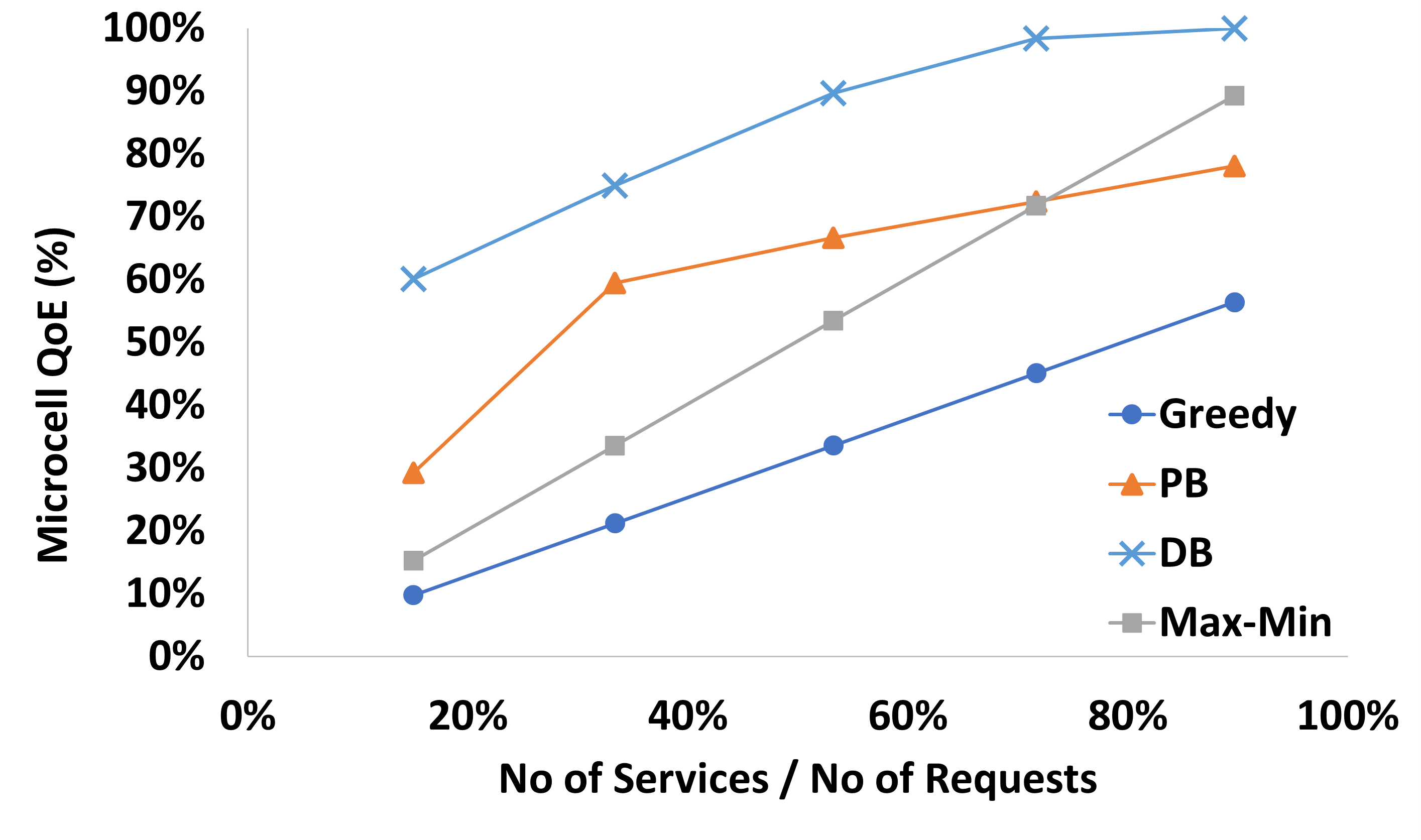}
\caption{The average of quality of experience}% in a microcell}
\label{fig:expMQoE}
\end{minipage}
\hfill
\begin{minipage}{.48\textwidth}
\centering
 \setlength{\abovecaptionskip}{0pt}
     \setlength{\belowcaptionskip}{-10pt}
\includegraphics[width=1\linewidth]{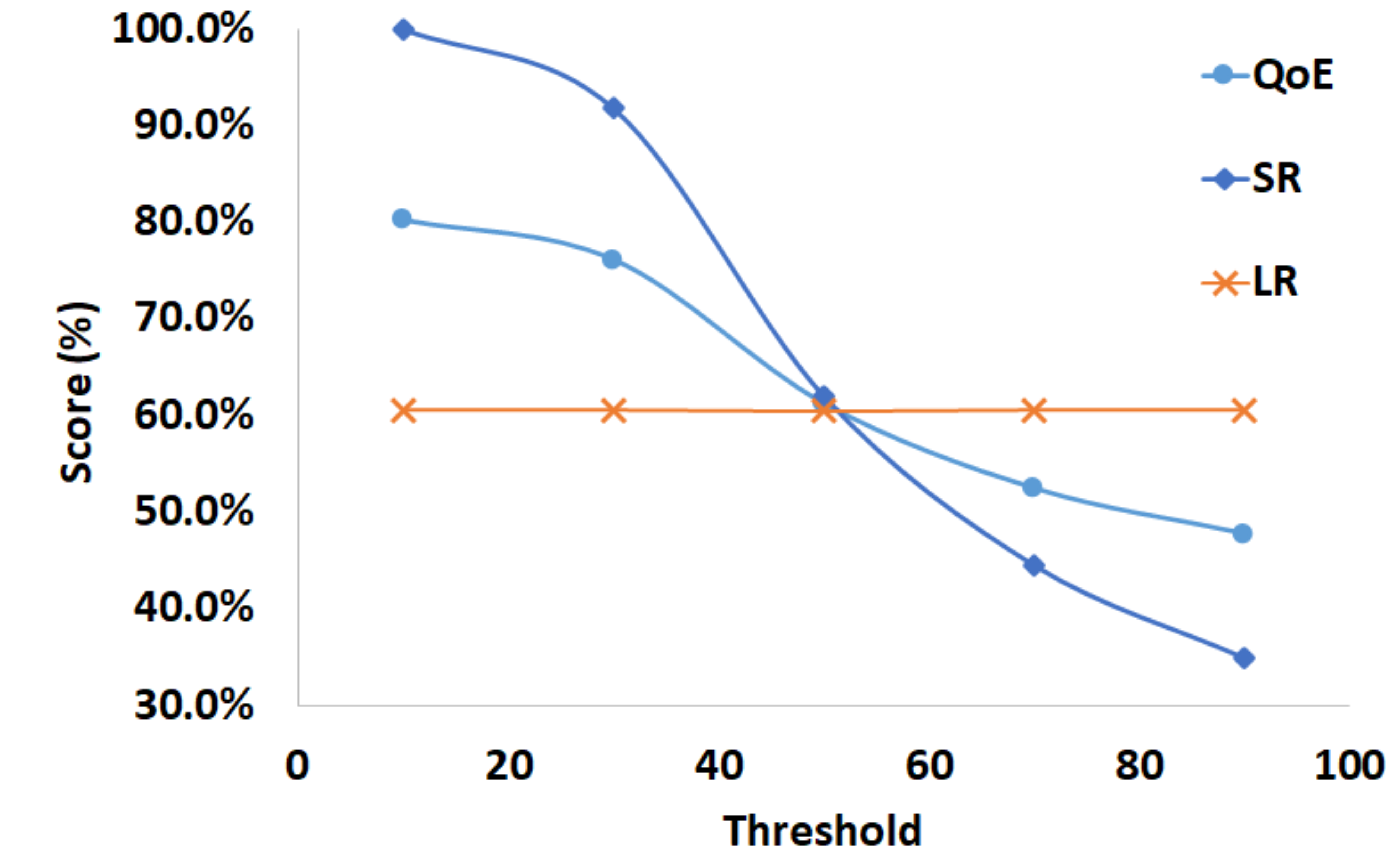}
\caption{The average of \{$QoE, \mathcal{SR}, \mathcal{FR}$\} using PB composition}% in a microcell}
\label{fig:expthreshold}
\end{minipage}
\end{figure}

\begin{figure}[!t]
\centering

\begin{minipage}{.48\textwidth}
\centering
 \setlength{\abovecaptionskip}{0pt}
     \setlength{\belowcaptionskip}{-4pt}
\includegraphics[width=1\linewidth]{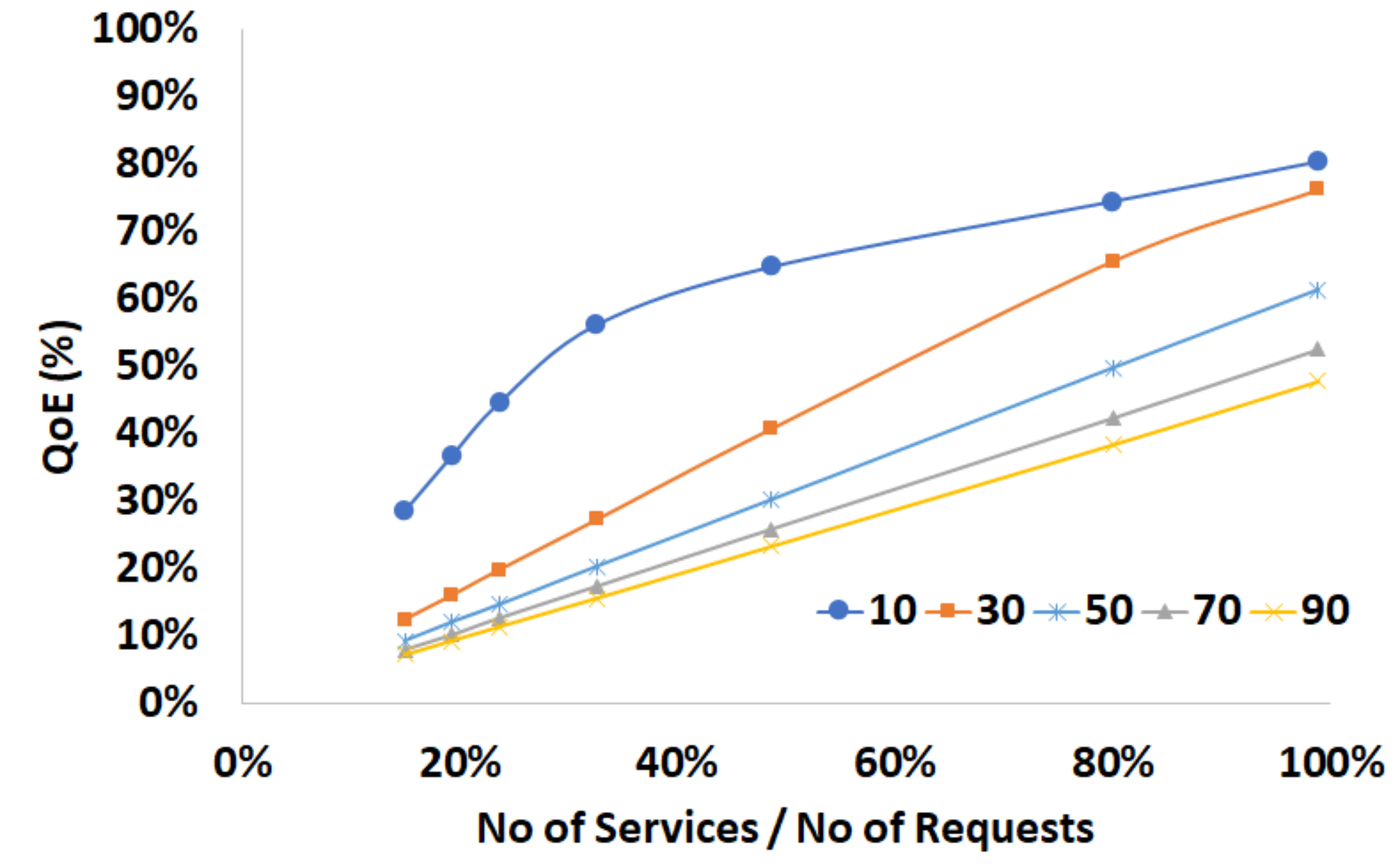}
\caption{The average of QoE using PB composition with various thresholds in a microcell }
%of \{10, 30, 50, 70, 90\}
\label{fig:expMoEthreshold}
\end{minipage}
\hfill
\begin{minipage}{.48\textwidth}
\centering
 \setlength{\abovecaptionskip}{0pt}
     \setlength{\belowcaptionskip}{-4pt}
\includegraphics[width=\linewidth]{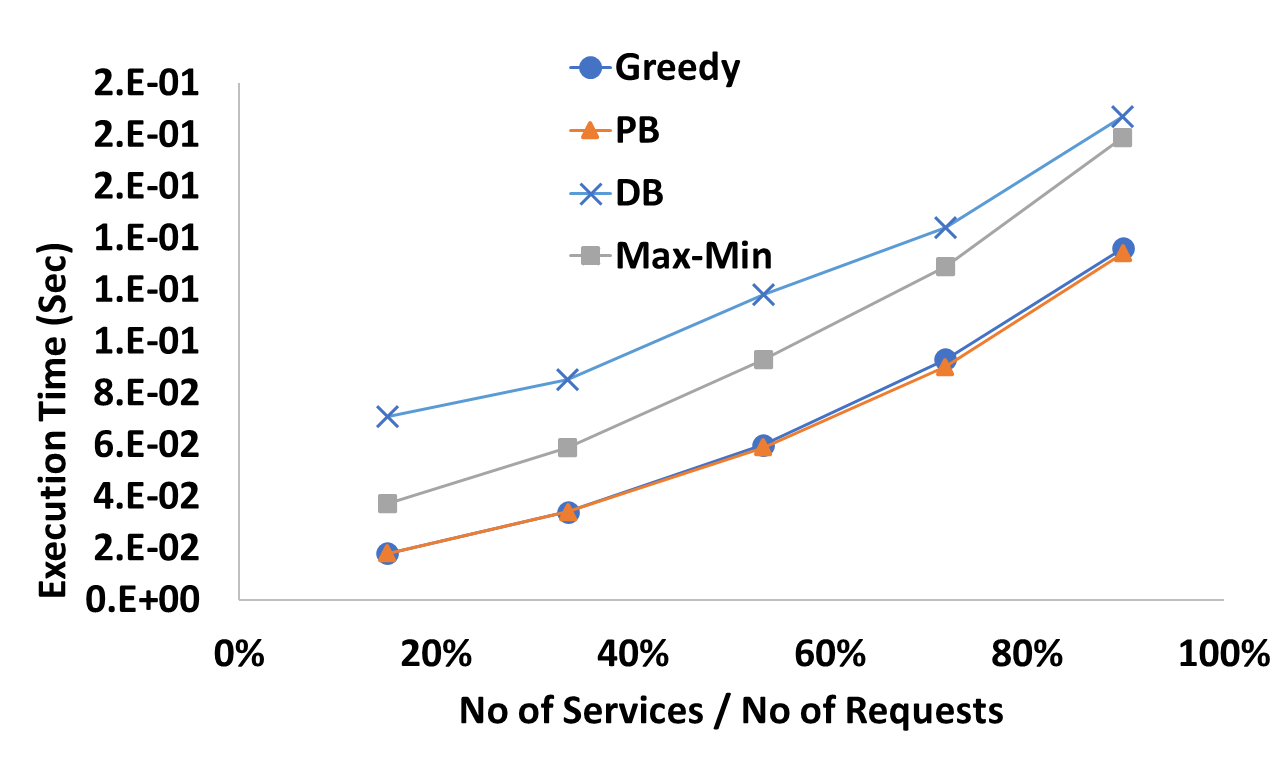}
\caption{The average execution time of all composition}
    \label{fig:compCost}
\end{minipage}

\end{figure}

Fig.\ref{fig:expthreshold} represents the impact of the threshold on the three previously tested attributes:  $\mathcal{SR}$, $\mathcal{FR}$, and QoE. We tested the PB approach with a 99\% ratio of services to requests. The x-axis in Fig.\ref{fig:expthreshold} {represents the threshold of partial services}. $FR$  does not change as the threshold increases, because it relies on the order of time slots and not the size of distribution (threshold) as discussed in the previous experiment. Also, both $\mathcal{SR}$ and QoE decrease as the threshold increases due to the thresholds' size. When the threshold's size increases, fewer consumers will be served. A lower number of fulfilled consumers results in low $\mathcal{SR}$ and thereby a low QoE.\looseness=-1

The fifth experiment compares the impact of the threshold on the QoE  with different ratios of services to requests.  We tested the PB approach with thresholds of \{10, 30, 50, 70, 90\}. %The x-axis in Fig.\ref{fig:expMoEthreshold} {represents the ratio of services to requests in the microcell}. 
In Fig.\ref{fig:expMoEthreshold},  the QoE increases when the number of available services increases for all threshold values. %  This observation can be explained, similar to the third experiment, by the availability of services to offer energy. The more services available, the more requests can be fulfilled. Therefore, the QoE of the microcell will increase. 
Additionally, the QoE for threshold 10 is the highest among all due to the threshold's size. When the size of the threshold is small, more consumers will be served. A higher number of fulfilled consumers results in high $\mathcal{SR}$ and thereby a high QoE. 
\vspace{-15pt}
\subsubsection{Computation Efficiency Evaluation}
%The seventh experiment evaluates the computation cost of the proposed approaches PB and DB, against Greedy. 
% Fig.\ref{fig:compCost} shows the average execution time of each approach.  
The execution time for all approaches increases with the increase in services' availability (See Fig.\ref{fig:compCost}). This is due to the increase in processing  time to assign these services.\looseness=-1

% \begin{figure}[!t]
%     \centering
%         \setlength{\abovecaptionskip}{-5pt}
%      \setlength{\belowcaptionskip}{-18pt}
%     \includegraphics[width=0.5\linewidth]{images/Exe.png}
%     \caption{The average execution time of all composition}
%     \label{fig:compCost}
% \end{figure}
\vspace{-15pt}
\section{Related Work}
\vspace{-8pt}
%The background of our work comes from two areas: energy sharing services and quality of experience. We present the related work to our research in both domains.
%\vspace{-5pt}
%\subsection{Energy Sharing Services}
%\vspace{-5pt}
Energy sharing services have been introduced as an alternative ubiquitous solution to charge IoT devices  \cite{lakhdari2020Vision}. %Energy service is defined as the wireless power transfer between IoT devices \cite{lakhdari2018crowdsourcing}.
Several studies have addressed challenges related to fulfilling the requirements of energy consumers \cite{lakhdari2018crowdsourcing}\cite{lakhdari2020Elastic}\cite{lakhdari2020fluid}. A temporal composition algorithm was proposed to compose energy services to fulfill a consumer's energy requirement \cite{lakhdari2018crowdsourcing}. The algorithm proposed the use of fractional knapsack to maximize the provided energy. An elastic composition was proposed to address the reliability of highly fluctuating energy providers \cite{lakhdari2020Elastic}. The composition uses the concepts of soft and hard deadlines to extend the stay of a consumer and select more reliable services. The intermittent behavior of energy services was addressed by a fluid approach \cite{lakhdari2020fluid}. The approach uses the mobility patterns of the crowd to predict the intermittent disconnections in energy services then replace or tolerate theses disconnections. Other studies tackled  challenges from a provider's perspective \cite{abusafia2020incentive}\cite{abusafia2020Reliability}. An context-aware incentive model was proposed to address the resistance in providing energy services \cite{abusafia2020incentive}. % The model used context-aware attributes to compute the rewards which encourage energy provision. 
Another article addresses the commitment of energy consumers to receive their initiated requests \cite{abusafia2020Reliability}. 
%The paper proposes a reliability model to evaluate consumers and compose the most reliable consumers for a single energy provider.
Existing literature in energy services addresses issues from a consumer or a provider perspective \cite{lakhdari2020Vision}. To the best of our knowledge, challenges related to the microcell perspective such as the QoE are yet to be addressed.

%\subsection{Quality of Experience}
Quality of experience (QoE) has several definitions in the literature based on the field of research \cite{fizza2021qoe}\cite{gong2009model}\cite{Amani2022QoE}. However, all existing definitions focus on assessing the quality of an application or a service based on the perception of the end-users. In addition, most of the literature focuses on assessing the QoE for multimedia applications. For instance, A method was proposed to gauge gaming QoE under system influencing factors such as delay, packet loss, and frame rates \cite{schmidt2020assessing}. % Another article proposed a model-based approach to measure the QoE for VoIP services. The model uses five quantifiable quality measures, including integrality, retainability, availability, usability, and instantaneousness \cite{gong2009model}.
Another study proposed ``Kaleidoscope" as an automated solution to evaluate Web features \cite{wang2019kaleidoscope}.  %Enhancing the QoE in terms of entertainment for restaurant customers using augmented reality was proposed \cite{margetis2013ieat}.
%, socialization, food selection, and ordering. 
As previously stated, the existing research focuses on assessing the QoE of a service based on the perception of the end-users. To the best of our knowledge, assessing the QoE in energy services is not explored yet. In addition, using energy services as a tool to enhance QoE in other microcell-based services is yet to be addressed. %\textit{This paper hence, is the first attempt to compose energy services with the objective of enhancing the QoE for consumers}.
\vspace{-15pt}
\section{Conclusion}
\vspace{-8pt}
We proposed an energy service composition framework that evaluates QoE in a microcell. A new QoE based-assessment was proposed to capture the overall satisfaction across consumers over a period of time. A two QoE-driven composition of energy service were proposed. The Partial-Based (PB) approach uses partial services to maximize the number of satisfied consumers and thereby increase the QoE. The Demand-Based (DB) approach uses partial services and prioritizes the most demanding time slots to maximize the number of satisfied consumers and their level of fulfillment and thereby increase the QoE. Experimental results show that DB outperforms all the evaluated approaches. The efficiency of the proposed approaches was investigated against a Greedy approach. Future direction is to consider the probability of change in the microcell energy demand.
\vspace{-25pt}
\section*{Acknowledgment}
\vspace{-10pt}
This research was partly made possible by  LE220100078 and LE180100158 grants from the Australian Research Council. The statements made herein are solely the responsibility of the authors.
\vspace{-10pt}

\bibliographystyle{unsrt}
\bibliography{main}

\begin{thebibliography}{10}

\bibitem{whitmore2015internet}
Andrew Whitmore, Anurag Agarwal, and Li~Da~Xu.
\newblock The internet of things—a survey of topics and trends.
\newblock {\em Information systems frontiers}, 17(2):261--274, 2015.

\bibitem{lakhdari2021fairness}
Abdallah Lakhdari and Athman Bouguettaya.
\newblock Fairness-aware crowdsourcing of iot energy services.
\newblock In {\em ICSOC}, pages 351--367. Springer, 2021.

\bibitem{lakhdari2018crowdsourcing}
Abdallah Lakhdari, Athman Bouguettaya, and Azadeh~Ghari Neiat.
\newblock Crowdsourcing energy as a service.
\newblock In {\em ICSOC}, pages 342--351, IEEE, 2018. Springer, Springer.

\bibitem{lakhdari2020composing}
Abdallah Lakhdari and et~al.
\newblock Composing energy services in a crowdsourced iot environment.
\newblock {\em IEEE TSC}, pages 1--1, 2020.

\bibitem{OvertheAirCharger}
Jessica Dolcourt.
\newblock Over-the-air wireless charging will come to smartphones, 2019.

\bibitem{muller1994expanded}
Christopher~C Muller and Robert~H Woods.
\newblock An expanded restaurant typology.
\newblock {\em Cornell Hotel and Restaurant Administration Quarterly},
  35(3):27--37, 1994.

\bibitem{chao2013c}
Tiffany Chao and et~al.
\newblock C-flow: Visualizing foot traffic and profit data to make informative
  decisions.
\newblock Technical report, University of Maryland, 2013.

\bibitem{fizza2021qoe}
Kaneez Fizza and et~al.
\newblock Qoe in iot: a vision, survey and future directions.
\newblock {\em Discover Internet of Things}, 1(1):1--14, 2021.

\bibitem{moller2014quality}
Sebastian M{\"o}ller and Alexander Raake.
\newblock {\em Quality of experience: advanced concepts, applications and
  methods}.
\newblock Springer, 2014.

\bibitem{wang2019kaleidoscope}
Pengfei Wang and et~al.
\newblock Kaleidoscope: A crowdsourcing testing tool for web quality of
  experience.
\newblock In {\em ICDCS}, pages 1971--1982. IEEE, 2019.

\bibitem{farris2010marketing}
Paul~W Farris and et~al.
\newblock {\em Marketing metrics: The definitive guide to measuring marketing
  performance}.
\newblock Pearson Education, 2010.

\bibitem{lakhdari2021proactive}
Abdallah Lakhdari and Athman Bouguettaya.
\newblock Proactive composition of mobile iot energy services.
\newblock In {\em ICWS}, pages 192--197. IEEE, 2021.

\bibitem{abusafia2020incentive}
Amani Abusafia, Athman Bouguettaya, and Sajib Mistry.
\newblock Incentive-based selection and composition of iot energy services.
\newblock In {\em IEEE SCC}, pages 304--311. IEEE, 2020.

\bibitem{kruse2007data}
Robert Kruse and et~al.
\newblock {\em Data structures and program design in C}.
\newblock Pearson, 2007.

\bibitem{huang2012predicting}
Ke~Huang and et~al.
\newblock Predicting mobile application usage using contextual information.
\newblock In {\em ubiquitous computing}, pages 1059--1065, 2012.

\bibitem{lakhdari2020Elastic}
Abdallah Lakhdari and et~al.
\newblock Elastic composition of crowdsourced iot energy services.
\newblock In {\em MobiQuitous}, pages 308--317, 2020.

\bibitem{yao2022wireless}
Jessica Yao and et~al.
\newblock Wireless iot energy sharing platform.
\newblock In {\em PerCom}, pages 118--120. IEEE, 2022.

\bibitem{lakhdari2020Vision}
Abdallah Lakhdari and et~al.
\newblock Crowdsharing wireless energy services.
\newblock In {\em IEEE CIC}, pages 18--24. IEEE, 2020.

\bibitem{lakhdari2020fluid}
Abdallah Lakhdari and Athman Bouguettaya.
\newblock Fluid composition of intermittent iot energy services.
\newblock In {\em SCC}, pages 329--336. IEEE, 2020.

\bibitem{abusafia2020Reliability}
Amani Abusafia and Athman Bouguettaya.
\newblock Reliability model for incentive-driven iot energy services.
\newblock In {\em MobiQuitous}, pages 196--205, 2020.

\bibitem{gong2009model}
Yan Gong and et~al.
\newblock Model-based approach to measuring quality of experience.
\newblock In {\em Inter. Conf. on Emerging Network Intelligence}, pages 29--32.
  IEEE, 2009.

\bibitem{Amani2022QoE}
Amani Abusafia and et~al.
\newblock Quality of experience optimization in iot energy services.
\newblock In {\em ICWS}. IEEE, 2022.

\bibitem{schmidt2020assessing}
Steven Schmidt and et~al.
\newblock Assessing interactive gaming quality of experience using a
  crowdsourcing approach.
\newblock In {\em QoMEX}, pages 1--6. IEEE, 2020.

\end{thebibliography}

\end{document}